\documentclass[aps,prb,showpacs,tightenlines, twocolumn,secnumarabic,nofootinbib,nobibnotes,superscriptaddress]{revtex4-1}
\usepackage{amsmath,amssymb,amsfonts,bm}
\usepackage{graphicx}
\usepackage{epstopdf}
\usepackage{dcolumn}
\usepackage{mathrsfs}  
\usepackage{appendix}
\usepackage[normalem]{ulem}
\usepackage[colorlinks=true,linkcolor=blue,citecolor=blue, urlcolor=blue,bookmarks=false]{hyperref}
\renewcommand{\Re}{\operatorname{Re}}
\renewcommand{\Im}{\operatorname{Im}}
\begin{document}
	\title{Tunable correlation effects of magnetic impurities by the cubic Rashba spin-orbit couplings}
	\author{Xiong-Tao Peng }
	\author{Fang Lin }
	\affiliation{
		Department of Physics, Ningbo University, Ningbo 315211, China}
	\author{Liang Chen}
	\affiliation{School of Mathematics and Physics, North China Electric Power University, Beijing, 102206, People’s Republic of China}
	\author{Lin Li} 
	\affiliation{
		College of Physics and Electronic Engineering, and Center for Computational Sciences,
		Sichuan Normal University, Chengdu 610068, China}
	\author{Dong-Hui Xu}
	\affiliation{Department of Physics and Chongqing Key Laboratory for Strongly Coupled Physics, Chongqing University, Chongqing 400044, China
	}
	\affiliation{Center of Quantum Materials and Devices, Chongqing University, Chongqing 400044, China}
	\author{Jin-Hua Sun}
	\email{sunjinhua@nbu.edu.cn}
	\affiliation{
		Department of Physics, Ningbo University, Ningbo 315211, China}

	\begin{abstract}

    We theoretically study the influence of the $k$-cubic Rashba spin-orbit coupling (SOC) on the correlation effects of magnetic impurities by combining the variational method and the Hirsch-Fye quantum Monte Carlo (HFQMC) simulations.
    Markedly different from the normal $k$-linear Rashba SOC, even a small cubic Rashba term can greatly alter the band structure and induce a Van Hove singularity in a wide range of energy, thus the single impurity local moment becomes largely tunable. 
    The cubic Rashba SOC adopted in this work breaks the rotational symmetry, but the host material is still invariant under the operations $\mathcal{R}^z(\pi)$, $\mathcal{IR}^z(\pi/2)$, $\mathcal{M}_{xz}$, $\mathcal{M}_{yz}$, where $\mathcal{R}^z(\theta)$ is the rotation of angle $\theta$ about the $z$-axis, $\mathcal{I}$ is the inversion operator and $\mathcal{M}_{xz}$ ($\mathcal{M}_{yz}$) is the mirror reflection about the $x$-$z$ ($y$-$z$) principal plane. 
    Saliently, various components of spin-spin correlation between the single magnetic impurity and the conduction electrons show three- or six-fold rotational symmetry.
    This unique feature is due to the triple winding of the spins with a $2\pi$ rotation of $\mathbf{k}$, which is a hallmark of the cubic Rashba effect, and can possibly be an identifier to distinguish the cubic Rashba SOC from the normal $k$-linear Rashba term in experiments. 
    Although the cubic Rashba term drastically alters the electronic properties of the host, we find that the spatial decay rate of the spin-spin correlation function remains essentially unchanged.  
    Moreover, the carrier-mediated Ruderman–Kittel–Kasuya–Yosida interactions between two magnetic impurities show twisted features, the ferromagnetic diagonal terms dominate when two magnetic impurities are very close, but the off-diagonal terms become important at long distances.

	\end{abstract}
	
	\maketitle

	\section{Introduction}
	Spin-orbit coupling (SOC) is a relativistic effect that locks the spin of a charge carrier with its angular momentum, and intense efforts have been made over the past decades to investigate and utilize SOCs in condensed matter physics. 
	There exist two representative SOCs, namely the Dresselhaus SOC caused by the bulk inversion asymmetry,\cite{Dresselhaus1955} and the Rashba SOC due to the spatial inversion asymmetry.\cite{Rashba1960,Uemura1974}
    In low-dimensional systems, the Rashba SOC becomes more important because it is stronger in the heterointerface,\cite{vasko1979,Rashba2d} and it is often described by the $k$-linear Rashba term, which can be written as $\propto (k_{-}\sigma_{+}-k_{+}\sigma_{-})$, where $k_{\pm}=k_x\pm ik_y $ denote the wave vectors and $\sigma_{\pm}=\sigma_x\pm i\sigma_y $ are the spin Pauli matrices.\cite{Bihlmayer_2015,manchon2015new,bihlmayer2022} 
    
	Besides the normal $k$-linear Rashba SOC, there also exists a higher-order term, namely the $k$-cubic Rashba SOC which has received continuous attention these years.\cite{gerchikov1992,winkler2003,Winkler2000,Zhao2020} The cubic Rashba SOC, which is often described by the Hamiltonian $\propto (k_{-}^3\sigma_{+}-k_{+}^3\sigma_{-})$, can greatly alter the dispersion relation and the effective field symmetry, and is predicted to induce larger spin Hall conductivity.\cite{John2005,Bleibaum2006,Matx2006} 
	The cubic Rashba SOC has been reported in a
	two-dimensional hole gas in inversion asymmetric
	semiconductors InGaAs and GaAs  heterostructures,\cite{Minkov2005,Noh2002} and a quasi- two-dimensional 
	electron gas formed at a
	surface of SrTiO$_3$ single crystal,\cite{Koga2012}
    and in rare-earth ternary materials TbRh$_2$Si$_2$\cite{Usachov2020prl} and EuIr$_2$Si$_2$.\cite{Usachov2020prb}
	
	As a prototypical strong correlation problem, Kondo effect in normal metals has
	been widely studied and well understood.\cite{anderson1961,kondo1964,Wilson1975} 
	The Kondo effect 
	is accompanied by the formation of Kondo cloud, which is characterized by the antiferromagnetic spin-spin correlation between the magnetic impurity and the conduction electrons. This spin-spin correlation function oscillates fast in space, and decays as $\sim 1/r^D$ when $r<\xi_K$, while it decays as $\sim 1/r^{D+1}$ if $r>\xi_K$,\cite{ishii1978,Barzykin1998,Borda2007} where $\xi_K$ is the Kondo length that extends to $\sim 1\ \mu m$ in typical metals,\cite{Moca2021} and has been confirmed recently via Fabry–P$\acute{e}$rot oscillations
	in conductance.\cite{Borzenets2020}
	
	The influence of $k$-linear Rashba SOC on the Kondo temperature $T_k$ has been studied previously using various methods, some indicate that $T_k$ is not significantly changed by Rashba SOC,\cite{malecki2007,zitko2011,Isaev2012} while others claim an exponential enhancement of $T_K$.\cite{Zarea2012,Chen_2018} Later, numerical renormalization group study found that for a fixed Fermi energy, the Kondo temperature $T_K$ varies weakly with Rashba SOC. 
%
	If instead, the band filling is low and held constant, increasing the Rashba SOC can drive the system into a helical regime where $T_K$ is exponentially enhanced.\cite{Wong2016} 
	Basically, one important reason to change the Kondo
	temperature is the divergence of density of states (DOS) which appears close to the band edge in the presence of $k$-linear Rashba SOC.\cite{Chen_2016} 
	On the other hand, in two-dimensional superconductors, it is found that $T_K$ is determined by the interplay between the Rashba SOC and superconducting energy gap, that the quantum phase transition between the magnetic doublet and Kondo singlet ground states is significantly affected by the Rashba SOC.\cite{lilin2018}  
	
	Moreover, the Kondo screening cloud shows anisotropy in both spatial and spin spaces in the presence of SOC.\cite{Feng2010,Feng_2011} It has also been proposed to use a magnetic impurity as a way to detect the Rashba effect through the local magnetization density of states. \cite{chirla2013}   
	Taking into account of the indirect exchange couplings between magnetic impurities, the Ruderman–Kittel–Kasuya–Yosida (RKKY)\cite{Kasuya1956,Kittel1954,Yosida1957} couplings become twisted in the presence of SOC.
	The RKKY interaction in two-dimensional systems with SOC can be written in a general form with three terms: Heisenberg, Ising, and Dzyaloshinskii-Moriya (DM) interactions, and this general form is valid for the Rashba SOC, the Dresselhaus SOC, and even when the two types of SOC are mixed.\cite{Hiroshi2004,Mross2009,Zhu2011} 
	   
	In this paper, we combine the variational method and the Hirsch-Fye quantum Monte Carlo (HFQMC)\cite{Hirsch1986} simulations to study the correlation effects of the impurities induced by the cubic Rashba SOC. 	
   The variational method has been widely used in the ground states of Anderson impurity problems in normal metals,\cite{Gunnarsson1983, Varma1976} systems with SOCs,\cite{Feng2010, Jinhua2015, Jinhua2018, Ma2018a, Wang2019, Yang2021} and superconductors.\cite{Simonin1995,Varma1999,Daniel2000,Huang_2022}
	The HFQMC technique is a numerically exact method which has been used to study magnetic impurities in metals, \cite{Hirsch1986,HFD1976,BU2008,Fye1988,Fye1987,Hirsch1987} dilute magnetic semiconductors,\cite{NK2007} graphene based systems\cite{HU2011,sun2013zigzag,sun2013magnetic,sun2018asymmetric} and in the presence of SOCs.\cite{sun2014,hufeiming2013}
	By combining the two methods, we can obtain not only a heuristic physical picture, but also the numerically exact results about the correlations.   
%
    The rest of the paper is organized as follows. In Sec.~\ref{Sec:Hamiltonian} we introduce the model Hamiltonian  and discuss the influence of the cubic Rashba term on the electronic properties of the host material. In Sec.~\ref{Sec:single_imp}, we show the results obtained using the variational method and the Hirsch-Fye quantum Monte Carlo simulations for single impurity case.
    The spin-spin correlation between two magnetic atoms, which is mediated by the conduction electrons, are given in Sec.~\ref{Sec:two_imp}.
     Finally, discussions and conclusions are given in Sec.~\ref{Sec:conclusion}.   
	
   \begin{figure}[t]
   	\begin{center}
   		\includegraphics[scale=0.10]{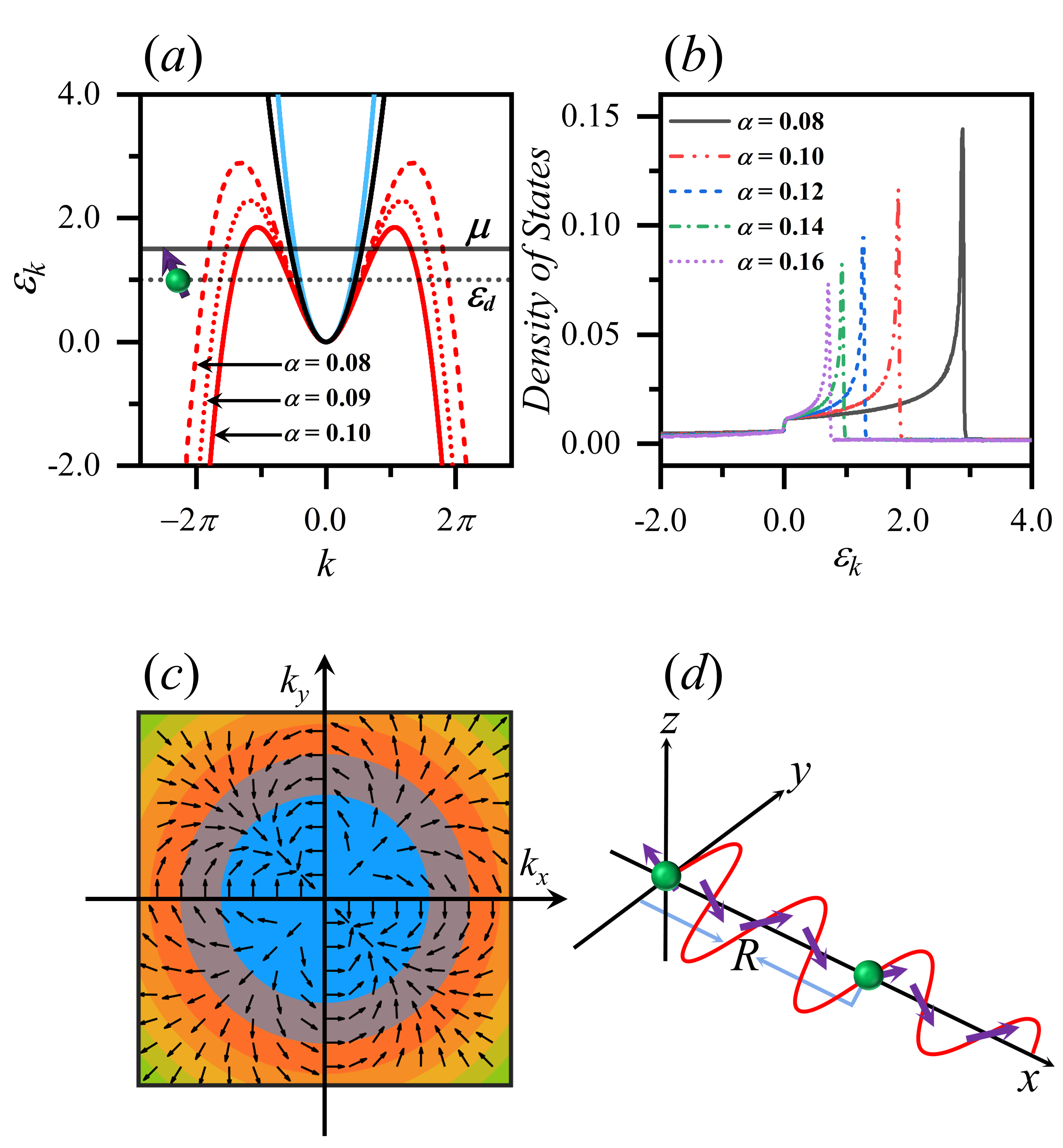}
   	\end{center}
   	\caption{\label{Fig1}(Color online) (a) The two-fold degenerate band (black line) splits into two bands due to the cubic Rashba SOC $\alpha$. The cubic Rashba term has marginal influence on the upper band (blue line). The red lines correspond to the lower bands with different $\alpha$ values, and they are drastically modified by $\alpha$. $\mu$ is the chemical potential, and $\varepsilon_d$ is the magnetic impurity energy level. (b) The density of states varies with $\alpha$, and Van Hove singularity emerges. (c) Spin texture caused by the cubic Rashba SOC in momentum space. (d)   
   		Schematic of the spin-spin interaction as a function of the distance
   		$R$ between two magnetic impurities. The red curve means the correlation strength, and the arrows denote the rotation of spin-spin interaction due to the cubic Rashba SOC.   
   	} 
   \end{figure}
	\section{Model Hamiltonian}\label{Sec:Hamiltonian}
    
    We use the Anderson impurity model to study the properties of magnetic impurities in a system with the cubic Rashba SOC term, 
    the total Hamiltonian is given by 
    \begin{equation}\label{Eq:Ham}
    \begin{aligned}
    H=H_0 + H_d + H_V.
    \end{aligned}
    \end{equation}
    $H_0$ describes the host material with the cubic Rashba SOC, $H_d$ is the magnetic impurity part, and $H_V$ denotes the hybridization between the local impurities and the conduction electrons. 
    The low-energy effective Hamiltonian of a host system with the cubic Rashba SOC is given by  
	\begin{equation}\label{Eq:H0}
	\begin{aligned}
	H_0=\sum_{\mathbf{k}}c_{\mathbf{k}}^\dagger\left[ h_0(\mathbf{k})-\mu \right] c_{\mathbf{k}},
    \end{aligned} 
	\end{equation}
	with
	\begin{equation}\label{Eq:h0detail}
	\begin{aligned}
    h_0(\mathbf{k})=\frac{\hbar^2\mathbf{k}^2}{2m}+\frac{i\alpha}{2}(k_-^3\sigma_+-k_+^3\sigma_-).
    \end{aligned}
	\end{equation}
	$h_0(\mathbf{k})$ is the single particle Hamiltonian incorporating cubic Rashba SOC,\cite{John2005,RM2014,zarea2006} and $c_{\mathbf{k}}^{\dagger}=(c_{\mathbf{k}\uparrow}^{\dagger},c_{\mathbf{k}\downarrow}^{\dagger})$ is the creation operator in spinor representation. The notations $k_{\pm}=k_x\pm ik_y$, $\sigma_{\pm}=\sigma_x\pm i\sigma_y$ are used to  denote the wave vectors and Pauli spin matrices. $\mu$ is the chemical potential, and $\alpha$ is the cubic Rashba term which can be adjusted experimentally. \cite{liu2018,cong2019,liuSY2005,KL2018} 
    Due to the SOC, the single particle eigenenergy splits from simple degenerate parabolic bands to two branches, 
    \begin{equation}\label{Eq:h0detail}
    	\begin{aligned}
    	\epsilon_{k\pm}=\frac{\hbar^2k^2}{2m} \pm \alpha k^3.
    	\end{aligned}
    \end{equation}

%

The magnetic impurity part is given by 
\begin{equation}\label{Eq:Hd}
\begin{aligned}
H_{d} =\sum _{ j , s=\uparrow,\downarrow }(\epsilon _d-\mu)d_{js}^{\dagger}d_{js} +\sum_{j}Ud_{j\uparrow }^{\dagger}d_{j\uparrow }d_{j\downarrow
}^{\dagger}d_{j\downarrow }.
\end{aligned}
\end{equation}
$j$ represents the magnetic impurity index. We study two cases, namely the single impurity doping and the two-impurity doping. When only one magnetic impurity is doped in the host, $j=1$. Otherwise if two impurities exist, $j=1,2$. 
$d_s^\dagger$ and $d_s$ are the creation and annihilation operators of the
spin-$s$ ($s=\uparrow, \downarrow$) state on the impurity site. $\epsilon_d$ is the impurity energy level which is beneath  $\mu$ in our calculations, and $U$ is the on-site Coulomb repulsion. 

Finally, the hybridization term between the localized state and the conduction electrons reads 
\begin{equation}\label{Eq:Hv}
\begin{aligned}
H_{V}=\sum _{\mathbf{k},j,\atop s=\uparrow,\downarrow }\left (e^{i\mathbf{k}\cdot\mathbf{R_j}}V_kd_{js}^\dagger c_{\mathbf{k}s }+H.c.\right).
\end{aligned}
\end{equation}
$V_k$ is the hybridization strength, and $\mathbf{R}_j$ is the coordinate of the $j$-th impurity. For two-impurity doping, we assume the two local atoms and conduction electrons have the same exchange coupling strength $V_k$ for simplification. 

In Fig.~\ref{Fig1}(a), we show the dispersion relation of the single particle energy bands given in Eq.~(\ref{Eq:h0detail}). The cubic Rashba SOC splits the degenerate parabolic band (black solid line) into two branches. One is $\epsilon_{k+}$ (blue solid line) and the other one is $\epsilon_{k-}$ (red lines). The cubic Rashba SOC term has minor effect on $\epsilon_{k+}$, but it can alter $\epsilon_{k-}$ significantly, as we can see from the three red lines, which correspond to $\epsilon_{k-}$ for slightly different $\alpha$ values. 
The DOS for different $\alpha$ values are plotted in Fig.~\ref{Fig1}(b). The cubic Rashba term largely modifies $\epsilon_{k-}$, consequently induce Van Hove singularity (VHS) into the host system, which is expected to greatly influences the local moment formation of magnetic impurities.  
Besides, the cubic Rashba SOC also breaks the rotational symmetry, but the system remains invariant under operations such as $\mathcal{R}^z(\pi)$, $\mathcal{IR}^z(\pi/2)$, $\mathcal{M}_{xz}$, $\mathcal{M}_{yz}$, where $\mathcal{R}^z(\theta)$ is the rotation of angle $\theta$ about the $z$-axis, $\mathcal{I}$ is the inversion operator and $\mathcal{M}_{xz}$ ($\mathcal{M}_{yz}$) is the mirror reflection about the $x$-$z$ ($y$-$z$) principal plane. The spin texture given in Fig.~\ref{Fig1}(c) reflects all these symmetries, which can be exhibited by the Kondo effect. Given in Fig.~\ref{Fig1}(d) is the schematic of our two-impurity case calculation. One magnetic impurity is fixed at the origin, and the other is located at a distance $R$ along the $x$-axis. 
In our calculations, the length unit is chosen as $k_0^{-1}$ which in typical metal is $k_0^{-1}\approx 10^{-9}m$. Correspondingly, the energy unit is $\frac{\hbar^2k_0^2}{2m^*}\approx 1.8\times 10^{-2}eV$ and the values of parameters $\alpha$, $U$, $\mu$, $V_k$, $\epsilon_d$ are given in units of $\frac{\hbar^2k_0^2}{2m^*}$.\cite{Feng_2011}
%
%
\section{Single impurity correlation effects}\label{Sec:single_imp}

	\subsection{The variational method}
 We can easily diagonalize $H_{0}$ and obtain a quasiparticle operator 
\begin{equation}\label{Eq:eigenstates}
	\begin{aligned}
		\gamma_{\mathbf{k}\pm}&=\dfrac{1}{\sqrt{2}}\left(e^{i\frac{3}{2}\theta_{\mathbf{k}}}c_{\mathbf{k}\uparrow}\pm ie^{-i\frac{3}{2}\theta_{\mathbf{k}}}c_{\mathbf{k}\downarrow} \right),
	\end{aligned} 
\end{equation}
where $\tan\theta_{\mathbf{k}}=k_{y}/k_{x}$, $\pm$ denotes the upper and lower energy bands. 
First let's discuss the simplest case when $H_{V}=0$ in which the magnetic impurity state decouples from the host material. Thus the ground state wave function of $H_{0}$ is given by
\begin{equation}
	\begin{aligned}
		|\Psi_0 \rangle =\prod _{\{\mathbf{k}\pm\}\in \Omega} \ \gamma _{\mathbf{k}\pm}^{\dagger}|0\rangle,
	\end{aligned}
\end{equation}
where $|0 \rangle$ is the vacuum, and the product runs over all the states within the Fermi sea $\Omega$. As for the impurity part, we assume that the Coulomb repulsion $U$ is large enough, and the impurity energy level $\epsilon_d$ is below the chemical potential $\mu$, so that the impurity site is always singly occupied by a local moment. The total energy of the system under this decoupled case is
\begin{equation}
	E_{0} = \epsilon_{d}-\mu+\sum_{\{\mathbf{k}\pm\}}\left(\epsilon_{\mathbf{k}\pm}-\mu \right).
\end{equation}
Then we consider the case with hybridization, where the trial wave function of the ground state is
\begin{equation}
	\begin{aligned}
		|\Psi \rangle =\left(a_{0}+\sum_{\{\mathbf{k}\pm\}}a_{\mathbf{k}\pm}d_{\mathbf{k}\pm}^{\dagger}\gamma_{\mathbf{k}\pm} \right)  |\Psi_0 \rangle,
	\end{aligned}\label{Eq:trialWF}	
\end{equation}
where $d_{\mathbf{k}\pm}=\dfrac{1}{\sqrt{2}}\left(e^{i\frac{3}{2}\theta_{\mathbf{k}}}d_{\uparrow}\pm ie^{-i\frac{3}{2}\theta_{\mathbf{k}}}d_{\downarrow} \right)$. $a_{0}$ and $a_{\mathbf{k}\pm}$
are variational parameters to be determined by optimizing the ground state energy.
The energy of the system in the trial state $|\Psi \rangle$ is given by
\begin{equation}
	\begin{aligned}
		E &= \dfrac{\langle \Psi|H|\Psi \rangle}{\langle \Psi|\Psi \rangle}\\
		&= \dfrac{\sum_{\{\mathbf{k}\pm\}}\left(E_{0}-\epsilon_{\mathbf{k}\pm} +\mu \right) a_{\mathbf{k}\pm}^{2}+2V_{k}a_{0}a_{\mathbf{k}\pm}+\left(\epsilon_{\mathbf{k}\pm}-\mu\right)a_{0}^{2}}{a_{0}^{2}+\sum_{\{\mathbf{k}\pm\}}a_{\mathbf{k}\pm}^{2}} \label{self-consistent}.
	\end{aligned}
\end{equation}
The variational method requires $\partial E/\partial a_{0}=\partial E/\partial a_{\mathbf{k}\pm}=0$, leading to 
\begin{equation}
	\begin{aligned}
		\left( E-\sum_{\{\mathbf{k}\pm\}}\left(\epsilon_{\mathbf{k}\pm}-\mu\right)\right)a_{0}&=\sum_{\{\mathbf{k}\pm\}}V_{k}a_{\mathbf{k}\pm},\\
		\left( E-E_{0}+\epsilon_{\mathbf{k}\pm}-\mu\right)a_{\mathbf{k}\pm}&= V_{k}a_{0}.
	\end{aligned}
\end{equation}
We can define the binding energy as $\Delta_{b}=E_{0}-E$, then the self-consistent equation is given by
\begin{equation}
	\begin{aligned}
		\left(\epsilon_{d}-\mu \right)=\Delta_{b}=\sum_{\{\mathbf{k}\pm\}}\dfrac{\left|V_{k}\right|^{2}  }{\epsilon_{\mathbf{k}\pm}-\mu-\Delta_{b}} .
	\end{aligned}
\end{equation}
If $\Delta_{b}>0$, the hybridized state is stable against the decoupled state. 
\begin{figure}[t]
	\includegraphics[scale=1.0]{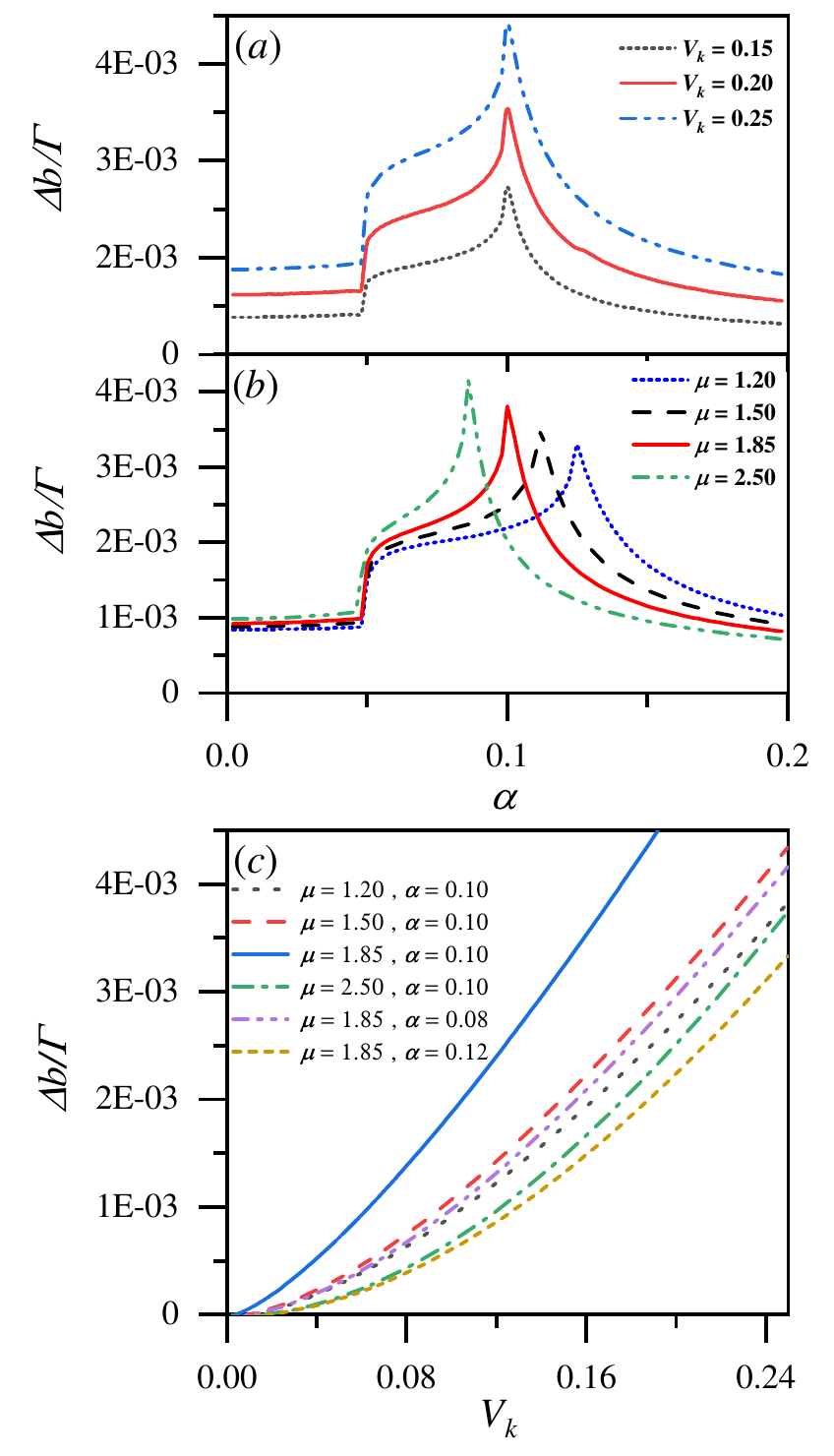}
	\caption{\label{0_Binding_energy}(Color online) Self-consistent results of the binding energy $\Delta_{b}$ for different combinations of parameter values. $\Delta_b$ versus $\alpha$ for (a) different $V_k$ when $\mu=1.85$ and (b) different $\mu$ when $V_{k}=0.20$. (c) $\Delta_b$ versus $V_k$ for various combinations of $\mu$, $\alpha$. $\Gamma$ is the energy cutoff chosen to be far away from $\mu$.
	} 
\end{figure}
In our variational method calculations, the impurity energy level is fixed slightly below the chemical potential, $\epsilon_{d}=\mu-0.001$, and the energy cutoff $\Gamma$ is chosen to be far away from $\mu$, that the low-energy physical properties will not be affected by the choice of $\Gamma$. 

We show the self-consistent results of the binding energy $\Delta_b$ for various combinations of $V_k$, $\mu$ and $\alpha$ in Fig.~\ref{0_Binding_energy}.
In Fig.~\ref{0_Binding_energy}(a) we show the binding energy with respect to the cubic Rashba term $\alpha$ for different values of $V_k$ when $\mu=1.85$. $\alpha$ greatly alters the band structure thus the DOS as is shown in  Fig.~\ref{Fig1}. We find that $\Delta_b$ shows a peak around $\alpha=0.1$. This is because for $\alpha=0.1$, the VHS occurs close to the chemical potential $\mu=1.85$. Besides, $\Delta_b$ is larger for stronger hybridization strength $V_k$, implying that the bound state is more easily formed for strong $V_k$. 
In Fig.~\ref{0_Binding_energy}(b), we fix $V_{k}=0.2$ and show similar results for various values of $\mu$. The energy corresponds to the VHS decrease monotonically with $\alpha$, and in a wide range of $\mu$ we can always observe the peak of $\Delta_b$. 
Fig.~\ref{0_Binding_energy}(c) shows the results of $\Delta_b$ versus $V_k$ for different combinations of $\mu$ and $\alpha$. In general, $\Delta_b$ is always positive due to the finite DOS in this system, and this is consistent with the previous results obtained using the same method.\cite{Feng2010,Jinhua2015,Jinhua2018} Larger values of $\Delta_b$ imply that the bound state is more stable. When $\alpha=0.10$, the VHS lies around $\mu=1.85$, such that the binding energy $\Delta_b$ (the blue straight line) becomes much larger than other cases.  
 
Next, we study the effect of the cubic Rashba SOC on the correlation between the local spin and the conduction electrons spins. This spin-spin correlation function measures the spatial Kondo screening cloud. The spin operator of the magnetic impurity spin is defined as 
$\mathbf{S}_{d}=\dfrac{1}{2}\sum_{s,s'}d_s^\dagger (\hat{\boldsymbol\sigma})_{s,s'} d_{s'}$ 
 and the conduction electron spin is
  $\mathbf{S}_{c}(\mathbf{r})=\dfrac{1}{2}\sum_{s,s'}c^{\dagger}_{s}(\mathbf{r} )(\hat{\boldsymbol{\sigma}})_{s,s'} c_{s'}( \mathbf{r})$, where $s,s'=\uparrow, \downarrow$.   
  By assuming the magnetic impurity location as the origin $\mathbf{r}=0$ and the conduction electron position as $\mathbf{r}$, the spin-spin correlation function is given by 
\begin{equation}
\begin{aligned}
J_{uv}\left( \mathbf{r}\right)=\langle S_{c}^{u}(\mathbf{r}) S_{d} ^{v}\left( 0\right) \rangle,  
\end{aligned}
\end{equation}
where $\langle \dots \rangle$ is the ground state average, and $u,v=x,y,z$ are the spin indices. $J_{uv}\left( \mathbf{r}\right)$ can be calculated by using the trial wave-function in Eq.~\ref{Eq:trialWF}. 
\begin{figure}[t]
	\includegraphics[width=8.8cm]{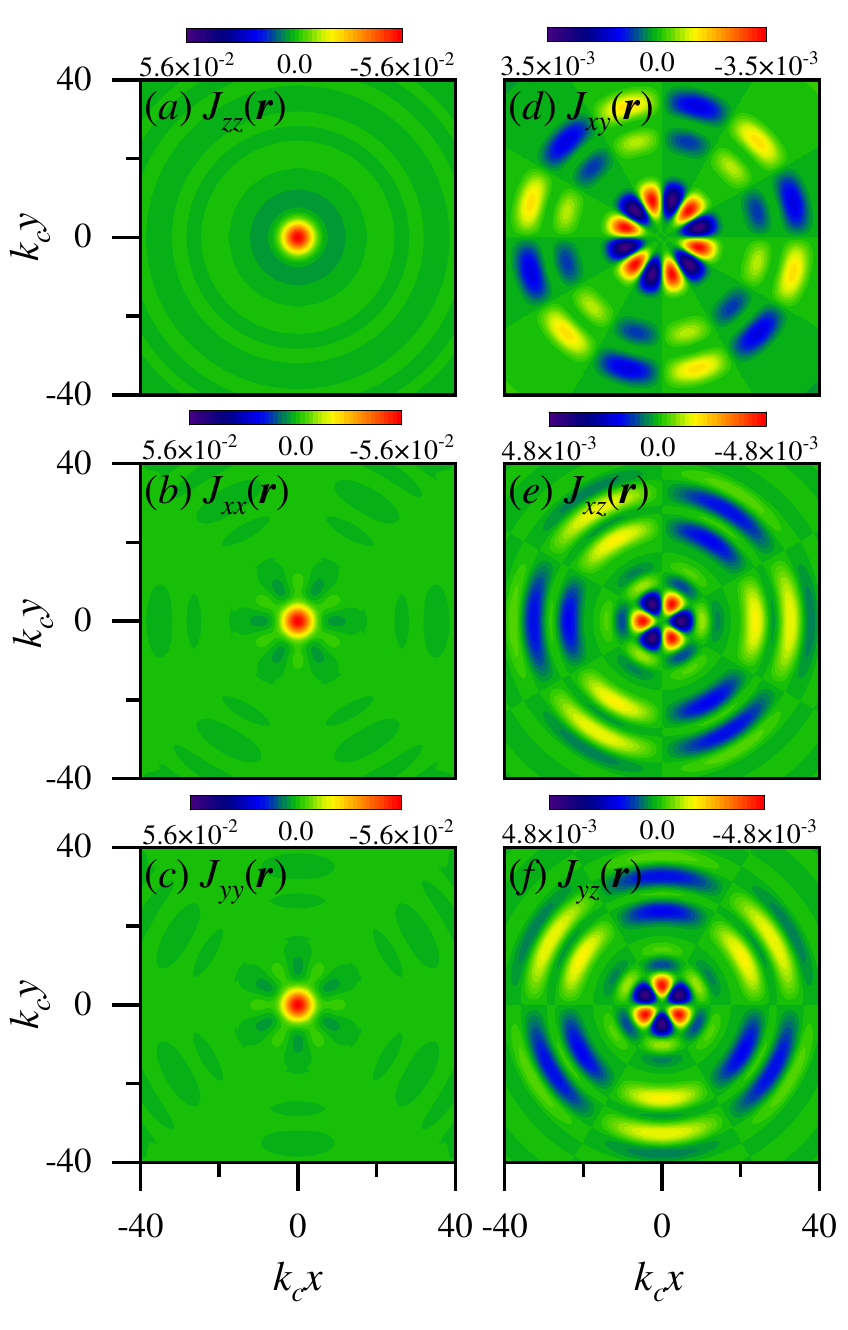}
	\caption{\label{0_2D}(Color online) The spatial pattern of the spin-spin correlation between magnetic impurity and conduction electrons with $(\alpha=0.1)$. (a)-(f) are for $J_{zz}(\mathbf{r})$, $J_{xx}(\mathbf{r})$, $J_{yy}(\mathbf{r})$, $J_{xy}(\mathbf{r})$, $J_{xz}(\mathbf{r})$, $J_{yz}(\mathbf{r})$. The parameters are $\Delta_{b}=0.02$, $\alpha =0.10$, $V_k=1.0$ and $\mu =1.50$. $k_{c}$ is the momentum cutoff chosen with respect to the energy cutoff $\Gamma$.} 
\end{figure}

\begin{figure}[htb]
	\includegraphics[scale=1]{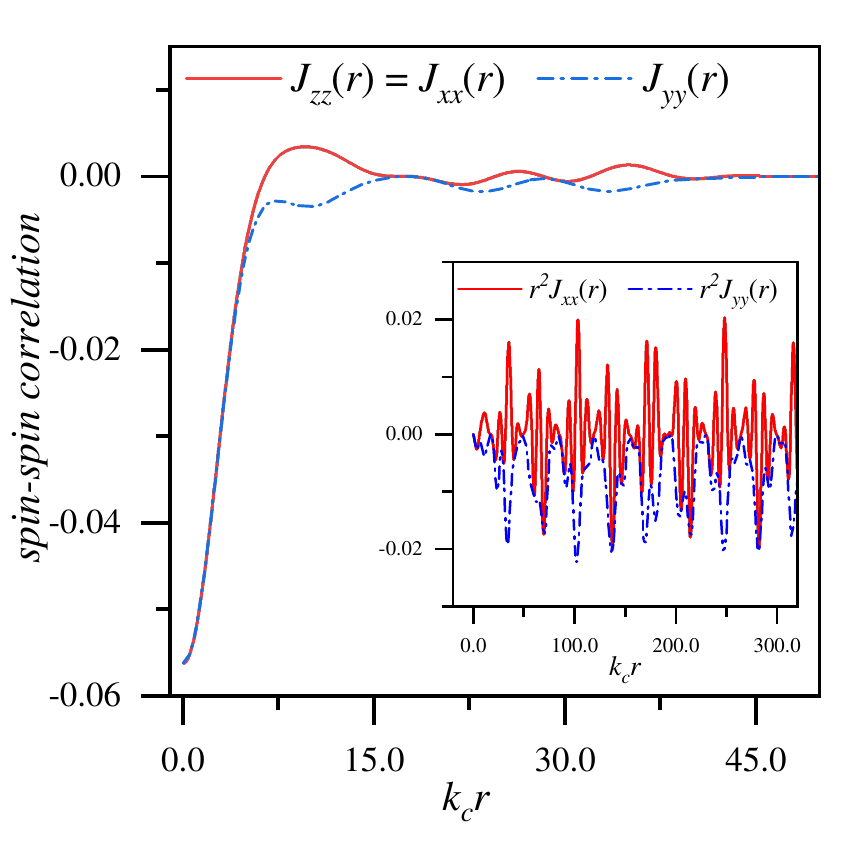}
	\caption{\label{0_1D}(Color online) Diagonal components of the Spin-spin correlation functions $J_{uu}\left( \mathbf{r}\right)$. The subfigure shows that the spin–spin correlation decays as $\propto 1/\mathbf{r}^{2}$ along the $x$-axis.  The parameters are the same as those in Fig. \ref{0_2D}. } 
\end{figure}

The diagonal and the off-diagonal terms take the form
\begin{equation}
	\begin{aligned}
		J_{zz}\left(\mathbf{r}\right)&=-\dfrac{1}{8}\left|B(\mathbf{r}) \right|^{2} + \dfrac{1}{8}\left|A(\mathbf{r}) \right|^{2},\\
		J_{xx}\left(\mathbf{r}\right)&=-\dfrac{1}{8}\left|B(\mathbf{r}) \right|^{2} - \dfrac{1}{8}\Re\left[ A(\mathbf{r}) \right] ^{2},\\
		J_{yy}\left(\mathbf{r}\right)&=-\dfrac{1}{8}\left|B(\mathbf{r}) \right|^{2} + \dfrac{1}{8}\Re\left[ A(\mathbf{r}) \right] ^{2},\\
		J_{xy}\left(\mathbf{r}\right)&=- \dfrac{1}{8}\Im\left[ A(\mathbf{r}) \right] ^{2},\\
		J_{xz}\left(\mathbf{r}\right)&=\dfrac{1}{4}\Im\left[ B(\mathbf{r})A(\mathbf{r}) \right], \\
		J_{yz}\left(\mathbf{r}\right)&=- \dfrac{1}{4}\Re\left[ B(\mathbf{r})A(\mathbf{r}) \right], \\
	\end{aligned}
\end{equation}
where $A(\mathbf{r})=\sum_{\{\mathbf{k}\pm\}}\pm e^{i( \mathbf{k}\cdot\mathbf{r}+3\theta_{\mathbf{k}}) }a_{\mathbf{k}\pm}$ and $B(\mathbf{r})=\sum_{\{\mathbf{k}\pm\}}e^{i\mathbf{k}\cdot\mathbf{r}}a_{\mathbf{k}\pm}$.
Due to the phase factors of $\gamma_{\mathbf{k}\pm}$ given in Eq.~\ref{Eq:eigenstates}, $A(\mathbf{r})$ contains the phase factor $3\theta_\mathbf{k}$ thus becomes three-fold rotational symmetric about the $z$-direction while $B(\mathbf{r})$ is isotropic in the $x$-$y$ plane.

In Fig.~\ref{0_2D}, we plot the spatial patterns of the spin-spin correlation function $J_{uv}(\mathbf{r})$ ($u,v=x,y,z$), and $k_{c}$ is the momentum cutoff chosen with respect to the energy cutoff $\Gamma$. $J_{zz}\left(\mathbf{r}\right)$ given in Fig.~\ref{0_2D}(a) is always isotropic about the origin while
 $J_{xx}\left(\mathbf{r}\right)$ and $J_{yy}\left(\mathbf{r}\right)$ given in Fig.~\ref{0_2D}(b), (c) are anisotropic because of the SOC in the $x$-$y$ plane. 
 We find that the diagonal components $J_{xx}\left(\mathbf{r}\right)$ and $J_{yy}\left(\mathbf{r}\right)$ have three-fold rotational symmetry about the $z$-direction. 
Note that the host system given in Eq.~\ref{Eq:H0} is not three-fold rotational symmetric. However, due to the phase factors of eigenstates given in Eq.~\ref{Eq:eigenstates}, the components of spin-spin correlation function show unique rotational symmetry. 
In addition, the host system is invariant under $\mathcal{R}^z(\pi)$, and consequently $J_{xx}\left(\mathbf{r}\right)$ and $J_{yy}\left(\mathbf{r}\right)$ also satisfy the six-fold rotational symmetry. 
All the diagonal terms are negative around $r=0$, indicating the antiferromagnetic coupling between the magnetic impurity spin and the conduction electron spins. 
The off-diagonal terms are merely induced by the SOC, and we find that 
$J_{xy}\left(\mathbf{r}\right) = J_{xy}\left(\mathbf{-r}\right)$ in Fig.~\ref{0_2D}(d), which can be analyzed using the $\mathcal{R}^z(\pi)$ symmetry of the host material. The other two off-diagonal components have the property $J_{xz}\left(x,y\right) = -J_{xz}\left(-x,y\right)$, and  $J_{yz}\left(x,y\right) = -J_{yz}\left(x,-y\right)$. 
Except for the isotropic $J_{zz}\left(\mathbf{r}\right)$, all the components of spin-spin correlation show either three- or six-fold rotational symmetry on the $x$-$y$ plane. 
The underlying reason for these unique symmetries is the 
triple winding of the spins with a complete $2\pi$ rotation of $\mathbf{k}$,\cite{Usachov2020prl, manchon2015new} which is a hallmark of the cubic Rashba effect, and can possibly be an identifier to distinguish the cubic Rashba SOC from the normal $k$-linear Rashba term in experiments.

All the components of the spin-spin correlation function oscillate and decay in space. To analyze the spatial decay rate of the correlations, in Fig.~\ref{0_1D} we show the diagonal components of the spin-spin correlation function along the $x$-axis. The parameters are chosen as $\Delta_{b}=0.02$, $\alpha =0.10$, $V_{k}=1.0$ and $\mu =1.50$. 
$J_{zz}\left(\mathbf{r}\right)=J_{xx}\left(\mathbf{r}\right)\neq J_{yy}\left(\mathbf{r}\right) $ along the $x$-axis. 
Shown in the subfigure is the results of $r^2J_{uu}(\mathbf{r})$ ($u=x,y$) along the $x$-axis. According to previous studies, the spin-spin correlation between the magnetic impurity and the conduction electrons decays as $\approx 1/r^{D+1}$ if $r>\xi_K$,\cite{ishii1978,Barzykin1998,Borda2007} where $\xi_K$ is the Kondo length. However, our variational calculations support
a $1/{r}^{2}$ decay for finite $\alpha$ at long distances. 
Even for the simple two-dimensional electron gas with $\alpha=0$, the decay rate of the spin-spin correlation function is still proportional to $1/{r}^{2}$ unless $\Delta_b>0.2$, which is unrealistically larger than the results of $\Delta_b$ obtained in Fig.~\ref{0_Binding_energy}.
We presume this is caused by the limitation of the variational method, and it is necessary to perform the unbiased HFQMC simulations to get more accurate results. 
   

\subsection{\label{sec:level2}Quantum Monte Carlo simulations}

The Hirsch-Fye algorithm naturally returns the imaginary-time Green's functions $g_{dd}^{ss'}(\tau)=- \langle T_\tau d_s(\tau)d_{s'}^{\dag}\rangle$, 
$G_{cd}^{ss'}(\mathbf{r},\tau)=- \langle T_\tau c_{\mathbf{r}s}(\tau)d_{s'}^{\dag}\rangle$, $G_{dc}^{ss'}(\mathbf{r},\tau)=- \langle T_\tau d_s(\tau)c_{\mathbf{r}s'}^{\dag}\rangle$, $G_{cc}^{ss'}(\mathbf{r},\tau)=- \langle T_\tau c_{\mathbf{r}s}(\tau)c_{\mathbf{r}s'}^{\dag}\rangle$,  where $s,s'=\uparrow,\downarrow$. 
In the HFQMC simulations, $\langle \cdots \rangle$ means taking the average over the discrete auxiliary
field. $\tau$ is the imaginary time ranges from 0 to $\beta$. All the information about the host material is included in the input non-interacting Green's functions $(U=0)$ which can be obtained analytically. 
\begin{figure}[t]
	\includegraphics[scale=1.0]{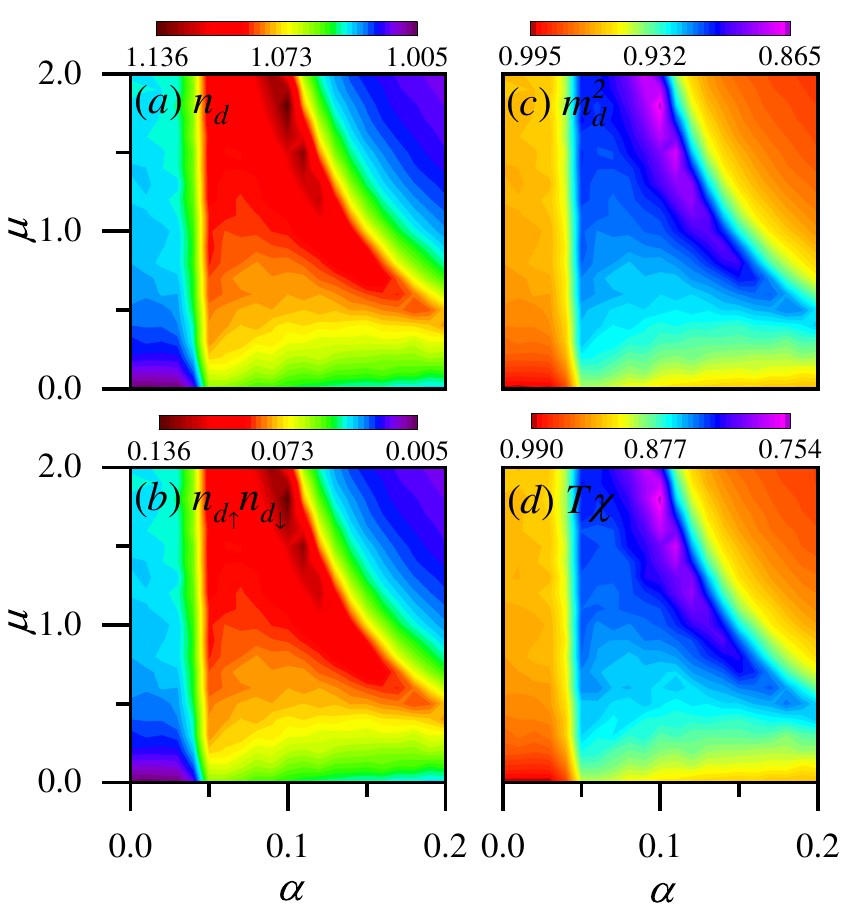}
	\caption{\label{0_local moment}(Color online) The HFQMC results of (a) $n_d$, (b) $n_{d_{\uparrow}}n_{d_{\downarrow}}$ , (c) $m_d^2$, and (d) $T\chi$ for various values of the chemical potential $\mu$ and the cubic Rashba term $\alpha$. We choose $U = 0.8$, $V_k=1.0$ and the temperature is $K_BT=1/32$.}
\end{figure}
By using the Green's function returned from the HFQMC simulations, we can calculate various quantities such as the expectation values of the total charge:
\begin{eqnarray*}
	n_d = \langle n_{d\uparrow}+n_{d\downarrow}\rangle,
\end{eqnarray*}
the local moment squared:
\begin{eqnarray*}
	m_d^2 = \langle (n_{d\uparrow}-n_{d\downarrow})^2\rangle,	
\end{eqnarray*}
the double occupancy:
\begin{eqnarray*}
	n_{d\uparrow\downarrow} = \langle n_{d\uparrow}n_{d\downarrow}\rangle,
\end{eqnarray*}
and the spin susceptibility:
\begin{eqnarray*}
	\chi = \int_0^\beta d\tau\langle (n_{d\uparrow}(\tau)-n_{d\downarrow}(\tau))(n_{d\uparrow}(0)-n_{d\downarrow}(0))\rangle.
\end{eqnarray*}
$\beta=1/k_BT$ is the inverse temperature.  
Note that the local moment squared on
the impurity site is given by $m_d^2=n_d-2n_{d\uparrow\downarrow}$, the closer this
value is to one, the more fully developed is the local moment.
In all our QMC simulations, we fix $\epsilon_d-\mu=-U/2$, namely the symmetric case in which the local moment formation is favored.\cite{anderson1961} 
\begin{figure}[t]
	\includegraphics[scale=1.0]{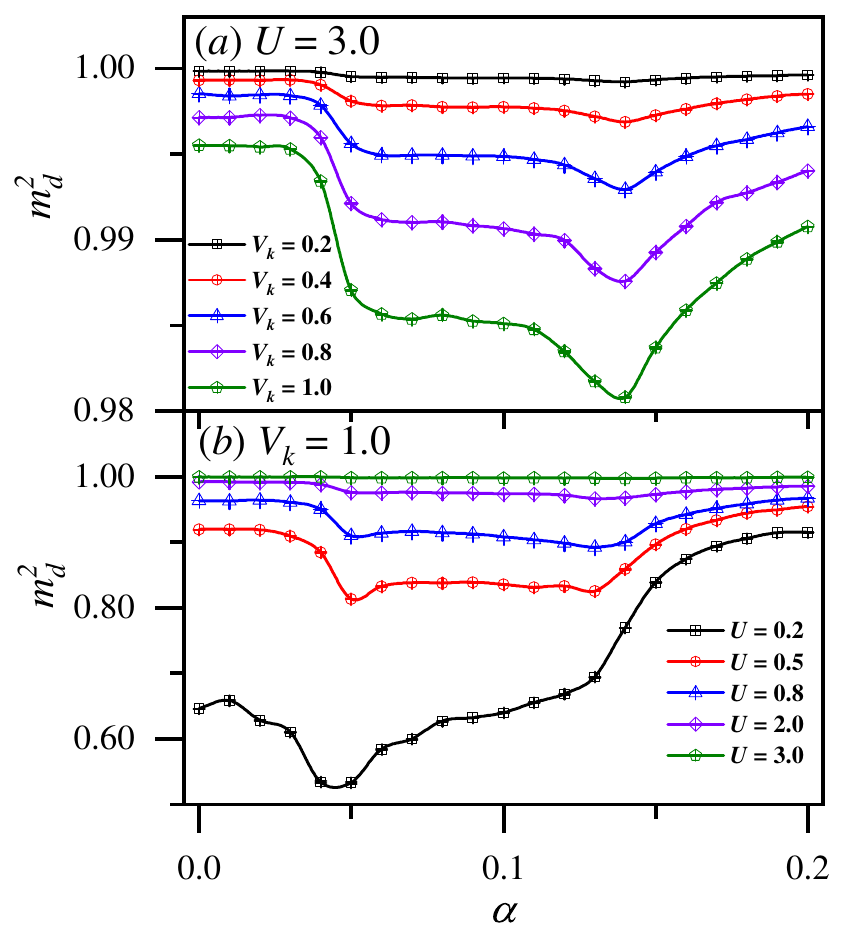}
	\caption{\label{fig6}(Color online) The HFQMC results of local moment squared for (a) $U=3.0$ with various $V_k$ values and (b) $V_k=1.0$ for different $U$ values. The chemical potential is
		$\mu=1.0$ and the temperature is $K_BT=1/32$.}
\end{figure}


In Fig.~\ref{0_local moment}, we show the thermodynamic quantities with respect to the chemical potential $\mu$ and the strength of the cubic Rashba term $\alpha$. The parameters are chosen as $U = 0.8$, $V_k=1.0$ and the
temperature is $K_BT=1/32$. The results for different parameter values shall remain qualitatively unchanged.  
 As is given in Fig.~\ref{Fig1}, small values of $\alpha$ can drastically modify the dispersion relation thus induce VHS. The energy corresponds to the VHS decreases as $\alpha$ increases, and this will influence the single magnetic impurity local moment. In Fig.~\ref{0_local moment}(a) we can see that the occupation number on the impurity site becomes larger in some regions, which corresponds to the case that the chemical potential is around the energy where VHS occurs. The double occupancy given in Fig.~\ref{0_local moment}(b) shows similar behavior, and the local moment is determined by the competition between the occupation and the double occupancy. 
 We can see that the local moment plotted in Fig.~\ref{0_local moment}(c) becomes smaller in the same region. It is natural that if the DOS at $\mu$ is large, the screening of the local magnetic impurity spin becomes stronger, so the local moment as well as the spin susceptibility shown in in Fig.~\ref{0_local moment}(d) are suppressed.  
%
%
%

In order to check the tunability of local moment by $\alpha$, we show the results of $m_d^2$ for different combinations of $U$ and $V_k$ in Fig.~\ref{fig6}. The chemical potential is fixed at
$\mu=1.0$ and the temperature is $K_BT=1/32$. In Fig.~\ref{fig6}(a) we choose $U=3.0$, and change the hybridization strength $V_k$. For all the values of $V_k$, we can find a dip of $m_d^2$ as we switch $\alpha$. The reduction of local moment is caused by the increase of DOS due to the cubic Rashba term $\alpha$. The change in $m_d^2$ is more obvious if $V_k$ is larger.   
In Fig.~\ref{fig6}(b), we show the local moment for different $U$ values while the hybridization is chosen as $V_k=1.0$. We can still see a dip of $m_d^2$ with as $\alpha$ varies, and the changes in local moment are more obvious for small $U$ values. In general, for a magnetic impurity with strong $V_k$ and relatively weak $U$, the local moment is largely tunable by switching the cubic Rashba SOC.

\begin{figure}[t]
	\includegraphics[scale=1.0]{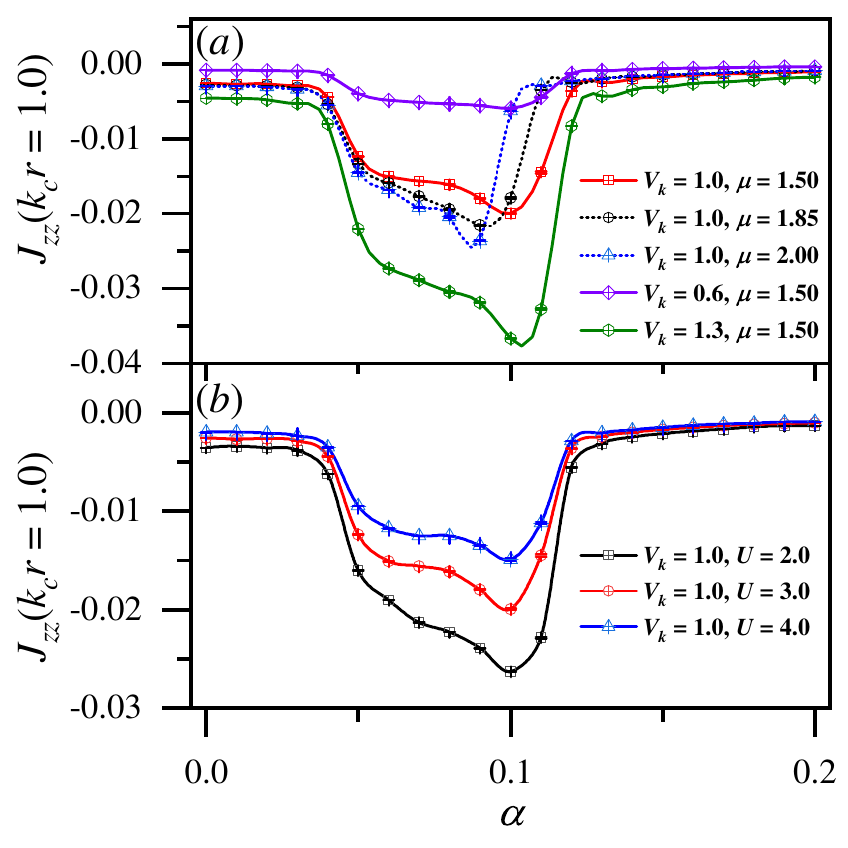}
	\caption{\label{0_ss-alpha}(Color online) The HFQMC results of $J_{zz}(\mathbf{r}=\{1.0,0\})$ with various combinations of $V_k$, $\mu$ and $U$. (a) The results of $J_{zz}(\mathbf{r})$ versus $\alpha$ for (a) fixed value of $U=3.0$, (b) for fixed $\mu=1.5$ and $V_k=1.0$. The
		temperature is chosen as $K_BT=1/16$.}
\end{figure}

In the following, the spin-spin correlation between the local magnetic impurity and the conduction electron is studied for different combinations of $\alpha$, $V_k$ and $U$ values. 
The spin-spin correlation between the magnetic impurity and conduction electron can be calculated from the Green's functions as\cite{gubernatis1987}:
\begin{eqnarray*}
	J_{zz}(\mathbf{r})=\langle S_d^z S_c^z \rangle &=&\langle (g_{dd}^{\uparrow\uparrow}-g_{dd}^{\downarrow\downarrow})\times(g_{cc}^{\uparrow\uparrow}-g_{cc}^{\downarrow\downarrow})-g_{dc}^{\uparrow\uparrow} \cdot g_{cd}^{\uparrow\uparrow}\notag\\
	&-&g_{dc}^{\downarrow\downarrow} \cdot g_{cd}^{\downarrow\downarrow}+g_{dc}^{\uparrow\downarrow} \cdot g_{cd}^{\downarrow\uparrow}+g_{dc}^{\downarrow\uparrow} \cdot g_{cd}^{\uparrow\downarrow}\rangle,
\end{eqnarray*}
\begin{eqnarray*}
		J_{xx}(\mathbf{r})=\langle S_d^x S_c^x \rangle &=&\langle(g_{dd}^{\uparrow\downarrow}+g_{dd}^{\downarrow\uparrow})\times(g_{cc}^{\uparrow\downarrow}+g_{cc}^{\downarrow\uparrow})-g_{dc}^{\uparrow\downarrow} \cdot g_{cd}^{\uparrow\downarrow}\notag\\
	&-&g_{dc}^{\downarrow\uparrow} \cdot g_{cd}^{\downarrow\uparrow}-g_{dc}^{\uparrow\uparrow} \cdot g_{cd}^{\downarrow\downarrow}-g_{dc}^{\downarrow\downarrow} \cdot g_{cd}^{\uparrow\uparrow}\rangle,
\end{eqnarray*}
\begin{eqnarray*}
		J_{yy}(\mathbf{r})=\langle S_d^y S_c^y \rangle &=&\langle-(g_{dd}^{\uparrow\downarrow}-g_{dd}^{\downarrow\uparrow})\times(g_{cc}^{\uparrow\downarrow}-g_{cc}^{\downarrow\uparrow})+g_{dc}^{\uparrow\downarrow} \cdot g_{cd}^{\uparrow\downarrow}\notag\\
	&+&g_{dc}^{\downarrow\uparrow} \cdot g_{cd}^{\downarrow\uparrow}-g_{dc}^{\uparrow\uparrow} \cdot g_{cd}^{\downarrow\downarrow}-g_{dc}^{\downarrow\downarrow} \cdot g_{cd}^{\uparrow\uparrow}\rangle,
\end{eqnarray*}
\begin{eqnarray*}
		J_{xy}(\mathbf{r})=\langle S_d^x S_c^y \rangle &=&\langle  i[(g_{dd}^{\uparrow\downarrow}+g_{dd}^{\downarrow\uparrow})\times(g_{cc}^{\uparrow\downarrow}-g_{cc}^{\downarrow\uparrow})-g_{dc}^{\uparrow\downarrow} \cdot g_{cd}^{\uparrow\downarrow}\notag\\
	&+&g_{dc}^{\downarrow\uparrow} \cdot g_{cd}^{\downarrow\uparrow}+g_{dc}^{\uparrow\uparrow} \cdot g_{cd}^{\downarrow\downarrow}-g_{dc}^{\downarrow\downarrow} \cdot g_{cd}^{\uparrow\uparrow}]\rangle,
\end{eqnarray*}
\begin{eqnarray*}
		J_{xz}(\mathbf{r})=\langle S_d^x S_c^z \rangle &=&\langle (g_{dd}^{\uparrow\downarrow}+g_{dd}^{\downarrow\uparrow})\times(g_{cc}^{\uparrow\uparrow}-g_{cc}^{\downarrow\downarrow})+g_{dc}^{\uparrow\downarrow} \cdot g_{cd}^{\downarrow\downarrow}\notag\\
	&-&g_{dc}^{\downarrow\uparrow} \cdot g_{cd}^{\uparrow\uparrow}-g_{dc}^{\uparrow\uparrow} \cdot g_{cd}^{\uparrow\downarrow}+g_{dc}^{\downarrow\downarrow} \cdot g_{cd}^{\downarrow\uparrow}\rangle,
\end{eqnarray*}
\begin{eqnarray*}
		J_{yz}(\mathbf{r})=\langle S_d^y S_c^z \rangle &=&\langle i[(g_{dd}^{\uparrow\downarrow}-g_{dd}^{\downarrow\uparrow})\times(g_{cc}^{\uparrow\uparrow}-g_{cc}^{\downarrow\downarrow})+g_{dc}^{\downarrow\uparrow} \cdot g_{cd}^{\uparrow\uparrow}\notag\\
	&+&g_{dc}^{\uparrow\downarrow} \cdot g_{cd}^{\downarrow\downarrow}-g_{dc}^{\uparrow\uparrow} \cdot g_{cd}^{\uparrow\downarrow}-g_{dc}^{\downarrow\downarrow} \cdot g_{cd}^{\downarrow\uparrow}]\rangle.
\end{eqnarray*}
We assume that the magnetic impurity is located at the origin of the coordinate, and $\mathbf{r}$ is the position of the conduction electron. 
In Fig.~\ref{0_ss-alpha} we show the results of $J_{zz}(\mathbf{r}=\{1.0,0\})$ with various combinations of $V_k$, $\mu$ and $U$. Given in Fig.~\ref{0_ss-alpha}(a) are the results of $J_{zz}(\mathbf{r})$ versus $\alpha$ for a fixed value of $U=3.0$. For all the parameters, we see that $J_{zz}(\mathbf{r})$ becomes stronger in a region as $\alpha$ increases. This region corresponds to the cases when the VHS emerges around the chemical potential $\mu$. In general, the values of $J_{zz}(\mathbf{r})$ grows with the hybridization strength $V_k$. In Fig. \ref{0_ss-alpha}(b), we fix the chemical potential as $\mu=1.5$, $V_k=1.0$, and present the results for different $U$. For all the values of $U$, we see similar behavior, that is the increase of $J_{zz}(\mathbf{r})$ in a certain region of $\alpha$. This indicates that the increase of spin-spin correlation is common for intermediate $U$ values.  Note that the relative magnitude of the $J_{zz}(\mathbf{r})$ is not always decrease monotonically with $U$, and it depends on the choice of $\mathbf{r}$. 
 
%

\begin{figure}[htb]
	\includegraphics[scale=1]{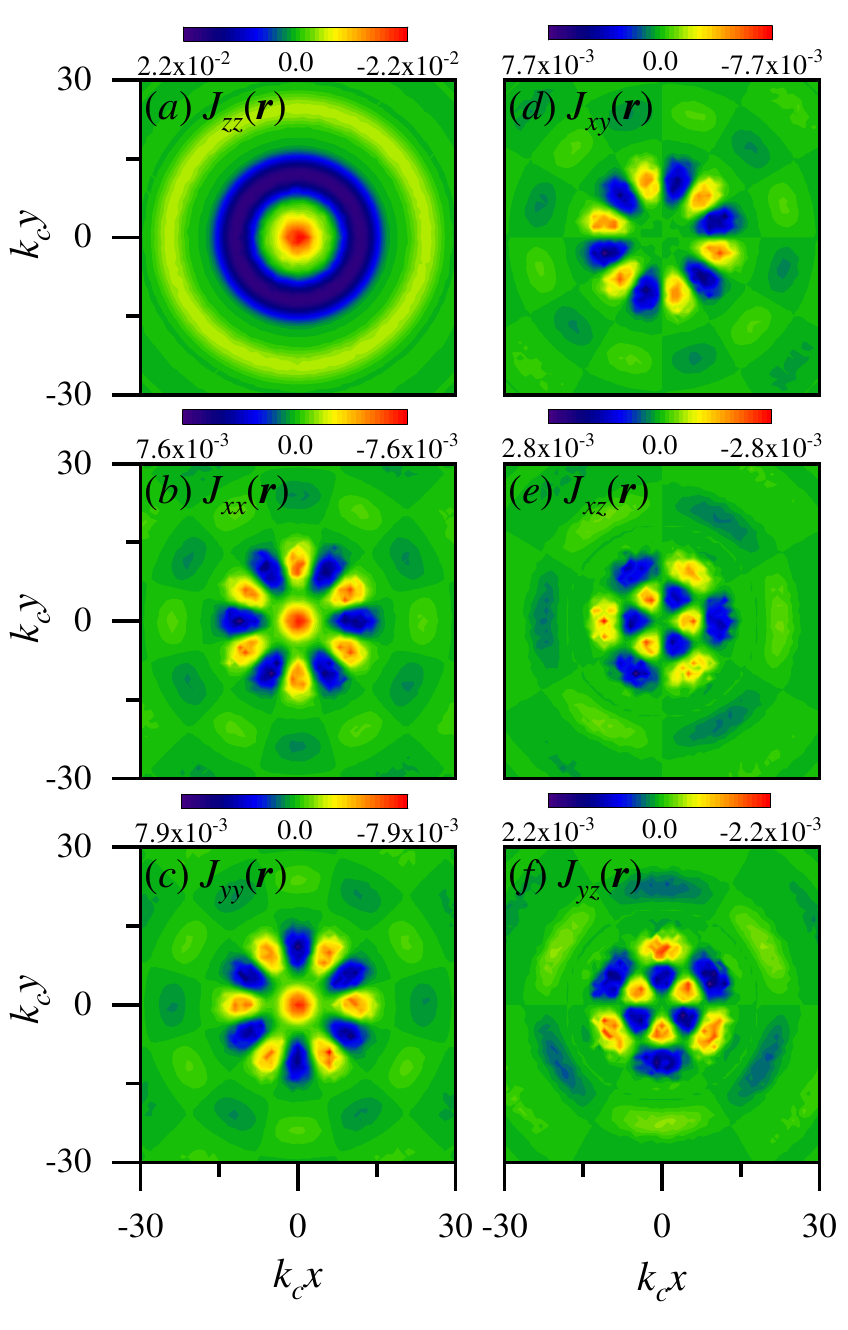}
	\caption{\label{0_SS2D}(Color online) The HFQMC results of the spin-spin correlation between the magnetic impurity and the conduction electrons in the $x$-$y$ plane.  (a)-(f) are for $J_{zz}(\mathbf{r})$, $J_{xx}(\mathbf{r})$, $J_{yy}(\mathbf{r})$, $J_{xy}(\mathbf{r})$, $J_{xz}(\mathbf{r})$, $J_{yz}(\mathbf{r})$.		
	The parameters are fixed as $\mu=1.5$, $\alpha=0.1$, $V_k=1.0$, $U=3.0$, and the temperature is $K_BT=1/16$.}
\end{figure}

In Fig.~\ref{0_SS2D} we show the HFQMC results of the spin-spin correlation between the magnetic impurity and the conduction electrons in the $x$-$y$ plane. The parameters are fixed as $\mu=1.5$, $\alpha=0.1$, $V_k=1.0$ and $U=3.0$, and $k_c$ is the momentum truncation. We can see that the HFQMC results of the spin-spin correlation exhibit basically the same symmetry with those obtained using the variational method as in Fig.~\ref{0_2D}. 
$J_{zz}(\mathbf{r})$ given in Fig.~\ref{0_SS2D}(a) is isotropic in the $x$-$y$ plane. The other two diagonal components $J_{xx}(\mathbf{r})$ and $J_{yy}(\mathbf{r})$ given in Fig.~\ref{0_SS2D}(b) and (c) are six-fold rotational symmetric, and $J_{xx}(\mathbf{r})=J_{yy}(\mathcal{R}^z(\frac{\pi}{2})\mathbf{r})$ which is consistent with the symmetry property of the host material, except for minor statistical errors caused in the QMC simulations. 
$J_{xy}(\mathbf{r})$ given in (d) generally has the same symmetry property with that obtained from the variational method.  Although $J_{xz}(\mathbf{r})$ and $J_{yz}(\mathbf{r})$ in (e) and (f) shows opposite signs in comparison to the counterparts in Fig.~\ref{0_2D}, all of them follows the three-fold rotational symmetry. 



\begin{figure}[htb]
	\includegraphics{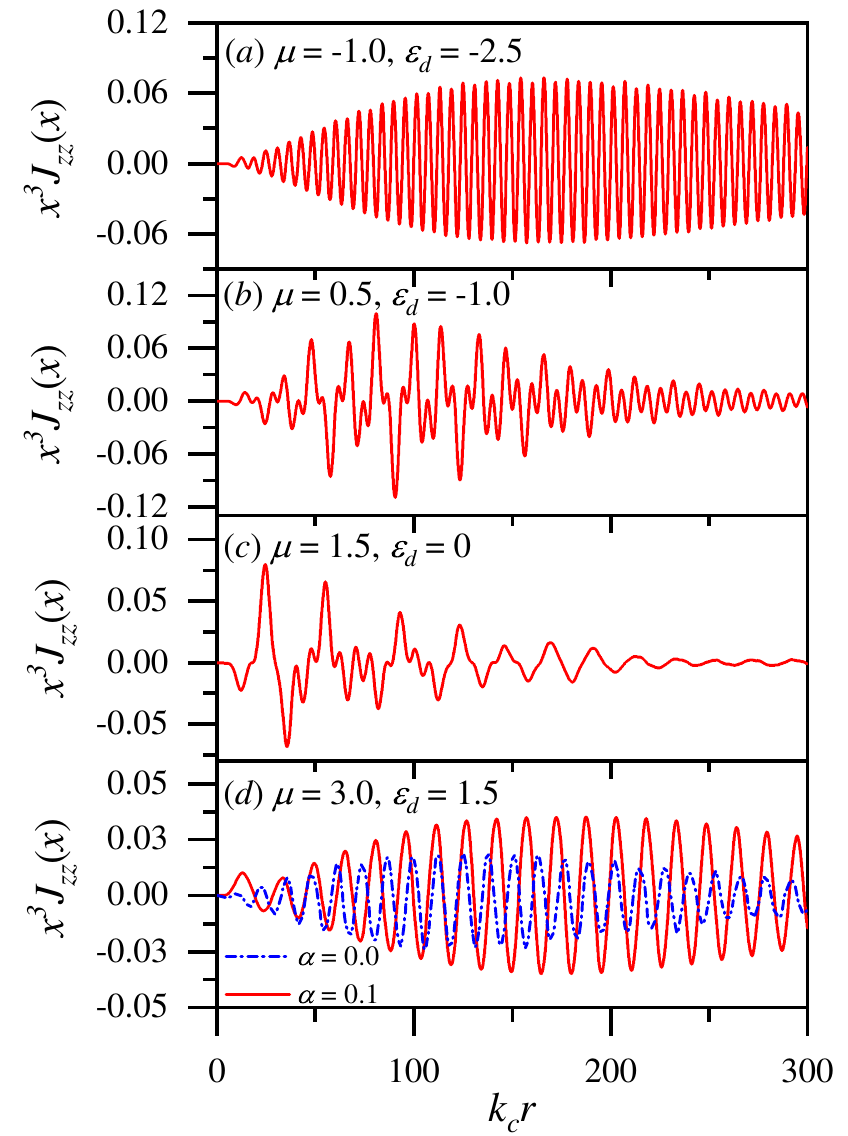}
	\caption{\label{0_shuaijian}(Color online) 
	The red solid lines show the results of $r^3J_{zz}(\mathbf{r})$ along the $x$-axis while $\alpha=0.1$ for (a) $\mu=-1.0$, (b) $\mu=0.5$, (c) $\mu=1.5$ and (d) $\mu=3.0$. We use the symmetric case $\epsilon_d-\mu=-U/2$ in our HFQMC simulations, so $\varepsilon_d$ varies with respect to $\mu$. The blue dashed line in (d) is the results for $10\times r^3J_{zz}(\mathbf{r})$ in a 2DEG with $\alpha=0$ for comparison. The parameters are  
	 $U=3.0$, $V_k=1.0$ and the temperature is $K_BT=1/16$.}
\end{figure}

In Fig.~\ref{0_shuaijian}, the red solid lines show the results of $r^3J_{zz}(\mathbf{r})$ along the $x$-axis while $\alpha=0.1$. The parameters are  
$U=3.0$, $V_k=1.0$ and the temperature is $K_BT=1/16$. 
When $\alpha=0.1$ and $\mu=-1.0$, as shown in Fig.~\ref{0_shuaijian}(a), only the lower band $\epsilon_{k-}$ involves in the screening process. We can see the spatial decay rate of the spin-spin correlation is about $r^{-3}$. As $\mu$ gradually increases, as in (b) and (c), both bands $\epsilon_{k-}$ and $\epsilon_{k+}$ take part in the Kondo screening, and the oscillation becomes more complicated.  
If $\mu=3.0$ as given in Fig.~\ref{0_shuaijian}(d), only the upper band $\epsilon_{k+}$ is responsible for the Kondo screening, and the decay rate of the $J_{zz}(x,0)$ is still proportional to $r^{-3}$, with different period of oscillation.
For comparison, the results of spin-spin correlation in a 2DEG for $\alpha=0$ is plotted as the blue dashed line in (d). Note that if $\alpha=0$, the spin-spin correlation is much smaller than the $\alpha=0.1$ case, so $r^3 J_{zz}(x,0)$ is multiplied by $10$ for clarity. 
Our results support the $1/r^3$ decay of the Kondo screening cloud at long distances, which is consistent with previous studies.\cite{ishii1978,Barzykin1998,Borda2007} 
We can see that the spatial decay rate obtained by the HFQMC results is more reliable than those given by the variational method shown in Fig.~\ref{0_1D}.  
Our HFQMC results show that the decay rate of the spin-spin correlation remains essentially unchanged in the presence of cubic Rashba term. However, the oscillation pattern and period are clearly affected by the cubic Rashba SOC. 

%
%

\section{Indirect coupling between TWO magnetic  IMPURIties}\label{Sec:two_imp}

Taking into account of the indirect coupling between two magnetic impurities, one natural question shall be how the RKKY interaction is influenced by the cubic Rashba term. For simplicity, we assume that one impurity is located at the origin, and the other impurity is on the $x$-axis with a distance $R$, as schematically plotted in Fig.~\ref{Fig1}(d). 
HFQMC returns the imaginary time Green's functions $G_{jj'}^{ss'}(\mathbf{R},\tau)$, where $j,j'=1,2$ mark the two magnetic atoms and $s,s'=\uparrow,\downarrow$ are the spin indices. 
The spin-spin correlation between two magnetic impurities measures the RKKY interaction mediated by the conduction electrons. 
The non-zero components of the spin-spin correlation function along the $x$-axis are\cite{sun2014}
\begin{eqnarray*}
	\langle S_1^z S_2^z \rangle&=&\langle S_1^x S_2^x \rangle \notag\\ &=&\langle (g_{11}^{\uparrow\uparrow}-g_{11}^{\downarrow\downarrow})\times(g_{22}^{\uparrow\uparrow}-g_{22}^{\downarrow\downarrow})-g_{12}^{\uparrow\uparrow} \cdot g_{21}^{\uparrow\uparrow}\notag\\
	&-&g_{12}^{\downarrow\downarrow} \cdot g_{21}^{\downarrow\downarrow}
	+g_{12}^{\uparrow\downarrow} \cdot g_{21}^{\downarrow\uparrow}+g_{12}^{\downarrow\uparrow} \cdot g_{21}^{\uparrow\downarrow}\rangle ,
\end{eqnarray*}
\begin{eqnarray*}
	\langle S_1^y S_2^y \rangle &=&\langle -(g_{11}^{\uparrow\downarrow}-g_{11}^{\downarrow\uparrow})\times(g_{22}^{\uparrow\downarrow}-g_{22}^{\downarrow\uparrow})+g_{12}^{\uparrow\downarrow} \cdot g_{21}^{\uparrow\downarrow}\notag\\
	&+&g_{12}^{\downarrow\uparrow} \cdot g_{21}^{\downarrow\uparrow}-g_{12}^{\uparrow\uparrow} \cdot g_{21}^{\downarrow\downarrow}-g_{12}^{\downarrow\downarrow} \cdot g_{21}^{\uparrow\uparrow}\rangle, \notag\\
\end{eqnarray*}
\begin{eqnarray*}
	\langle S_1^x S_2^z \rangle &=&-\langle S_1^z S_2^x \rangle\\
	&=&\langle (g_{11}^{\uparrow\downarrow}+g_{11}^{\downarrow\uparrow})\times(g_{22}^{\uparrow\uparrow}-g_{22}^{\downarrow\downarrow})+g_{12}^{\uparrow\downarrow} \cdot g_{21}^{\downarrow\downarrow}\notag\\
	&-&g_{12}^{\downarrow\uparrow} \cdot g_{21}^{\uparrow\uparrow}-g_{12}^{\uparrow\uparrow} \cdot g_{21}^{\uparrow\downarrow}+g_{12}^{\downarrow\downarrow} \cdot g_{21}^{\downarrow\uparrow}\rangle. \notag\\
\end{eqnarray*}
\begin{figure}[t]
	\begin{center}
		\includegraphics[scale=1]{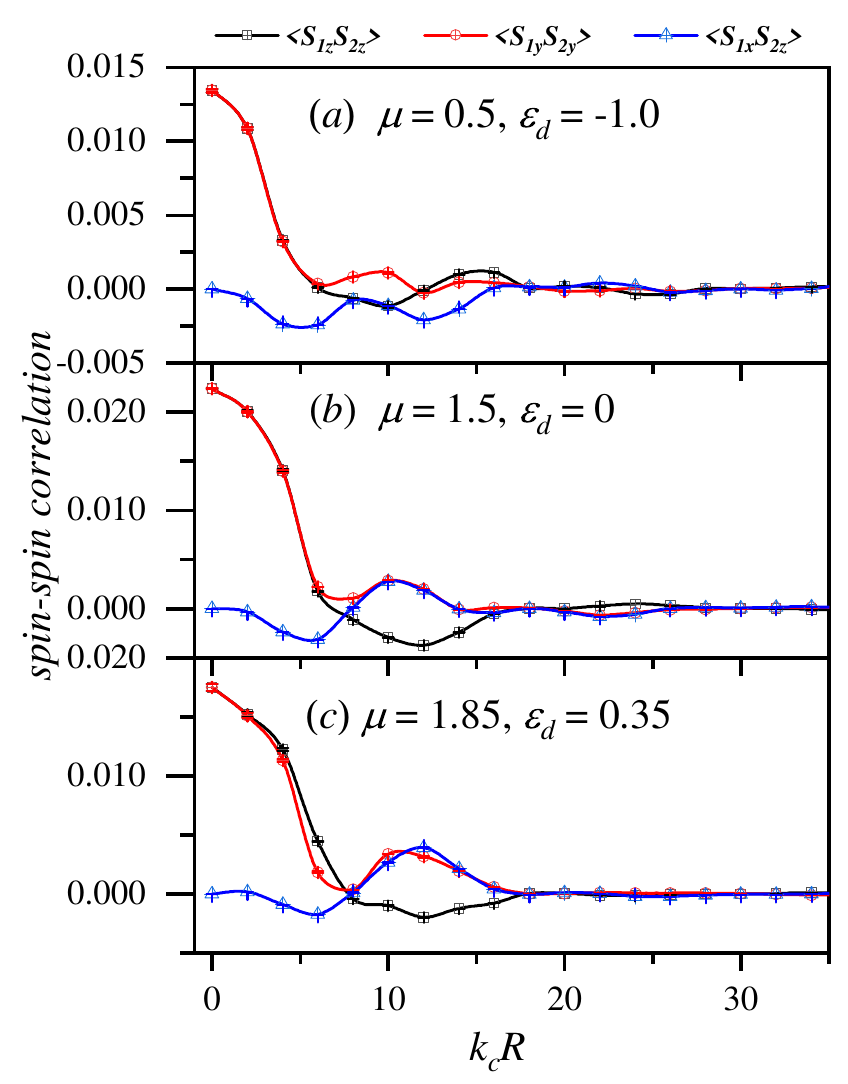}
	\end{center}
	\caption{\label{0_rkky}(Color online) The spin-spin correlation between two magnetic impurities with respect to the distance between them $R$. (a)-(c) are listed in the order of increasing $\mu$, and $\epsilon_d-\mu=-U/2$. $k_c$ is the momentum truncation, and the parameters are chosen as $U=3.0$ and $V_k=1.0$, $k_BT=1/8$. The cubic Rashba SOC term is $\alpha=0.1$, and VHS occurs at energy $1.85$. 
	}
\end{figure} 
In Fig.~\ref{0_rkky} we show the spin-spin correlation between the two magnetic impurities with respect to the distance $R$ between them, and $k_c$ is the momentum truncation. The parameters are chosen as $\alpha=0.1$, $U=3.0$ and $V_k=1.0$, $k_BT=1/8$. 
We consider the symmetric case, with $\epsilon_d-\mu=-U/2$. Along the $x$-axis, we can see that $\langle S_1^z S_2^z \rangle=\langle S_1^x S_2^x \rangle \neq \langle S_1^y S_2^y \rangle$. This is due to the cubic Rashba term $\alpha$, without which all the three components shall be exactly the same. When $\alpha=0.1$, the VHS emerges at energy value $\mu=1.85$. 
Figs.~\ref{0_rkky}(a)-(c) are listed in the order of increasing $\mu$. 
$\mu=0.5$ given in Fig.~\ref{0_rkky}(a) corresponds to relatively low DOS, while $\mu=1.50$ and $\mu=1.85$ given in Figs.~\ref{0_rkky}(b) and (c) are close to the energies where VHS occurs. 
We can see that the diagonal terms $\langle S_1^z S_2^z \rangle$ and $\langle S_1^y S_2^y \rangle$ are suppressed when $\mu=0.5$. As $\mu$ approaches the VHS point, the DOS increases, and so the diagonal terms $\langle S_1^z S_2^z \rangle$ and $\langle S_1^y S_2^y \rangle$.   
For all the cases, the diagonal terms are positive and dominant when the two impurities are close, indicating that the two magnetic impurities are ferromagnetically correlated, and the values oscillate and decay in space. The only non-zero off-diagonal term $\langle S_1^x S_2^z \rangle=-\langle S_1^z S_2^x \rangle$ also changes with respect to the values of $\mu$. The off-diagonal terms correspond to the DM interaction,\cite{Hiroshi2004,Mross2009,Zhu2011} and it is a manifestation of the SOC in the host material. 
At distance $k_cR\sim 10$, the off-diagonal correlation $\langle S_1^x S_2^z \rangle$ is of the same order of magnitude as the diagonal terms. 

%

\section{conclusions}\label{Sec:conclusion}   
In this paper, we apply the variational method and the HFQMC technique to study the influence of the $k$-cubic Rashba SOC on the correlation effects of magnetic impurities. 
The cubic Rashba SOC greatly alters the band structure and induces a VHS to the host material. The $k$-linear Rashba SOC can also cause the divergence of DOS, but the divergence occurs at the bottom of the bands. However, the VHS induced by the cubic Rashba SOC occurs in a very wide range of energy, and the single impurity local moment becomes largely tunable, especially for strong $V_k$ and relatively weak $U$. 
Both the variational method and the HFQMC simulations support the three- or six-fold rotational symmetry of the various components of spatial spin-spin correlation. 
This unique feature is a manifestation of the cubic Rashba SOC,  and can possibly be used in experiments to distinguish the cubic Rashba SOC from the normal $k$-linear Rashba term.  
The HFQMC calculations show that the $1/r^3$ decay rate of this spin-spin correlation is essentially unchanged by the cubic Rashba SOC term $\alpha$. 
Moreover, the RKKY couplings between two magnetic impurities displays very complicated form. Besides the normal diagonal components,
we still obtain the finite off-diagonal components, which corresponds to the DM interaction between two magnetic impurities, and they become the same order of magnitude as the diagonal terms at distance $k_cR\approx 10$.  
%
%

\section{Acknowledgments}
J.-H.S. acknowledges financial support from the Zhejiang Provincial Natural Science Foundation of China (Grant No. LY19A040003) and K.C.Wong Magna Fund in Ningbo University. D.-H.X. was supported by the NSFC (under Grant Nos. 12074108 and 12147102) and the Natural Science Foundation of Chongqing (Grant No. CSTB2022NSCQ-MSX0568). L.C. was supported by the NSFC ( under Grant No. 12174101) and the Fundamental Research Funds for the Central Universities (Grant No. 2022MS051)

\newpage
\bibliographystyle{apsrev4-1}
\bibliography{ref}

%
%
%
%
%
%
%
%
%
%
%
%
%

\end{document}